# Terahertz phase slips in striped La$_{2-x}$Ba$_x$CuO$_4$

D. Fu[1], D. Nicoletti[1], M. Fechner[1], M. Buzzi[1], G. D. Gu[2], and A. Cavalleri[1,3]*

[1] Max Planck Institute for the Structure and Dynamics of Matter, 22761 Hamburg, Germany

[2] Condensed Matter Physics and Materials Science Department, Brookhaven National Laboratory, Upton, NY, United States

[3] Department of Physics, Clarendon Laboratory, University of Oxford, Oxford OX1 3PU, United Kingdom

* e-mail: andrea.cavalleri@mpsd.mpg.de

## Abstract

**Interlayer transport in high-$T_C$ cuprates is mediated by superconducting tunneling across the CuO$_2$ planes. For this reason, the terahertz frequency optical response is dominated by one or more Josephson plasma resonances and becomes highly nonlinear at fields for which the tunneling supercurrents approach their critical value, $I_C$. These large terahertz nonlinearities are in fact a hallmark of superconducting transport. Surprisingly, however, they have been documented in La$_{2-x}$Ba$_x$CuO$_4$ also above $T_C$ for doping values near $x$ =1/8, and interpreted as an indication of superfluidity in the stripe phase. Here, Electric Field Induced Second Harmonic (EFISH) is used to study the dynamics of time-dependent interlayer voltages when La$_{2-x}$Ba$_x$CuO$_4$ is driven with large-amplitude terahertz pulses, in search of other characteristic signatures of Josephson tunnelling in the normal state. We show that this method is sensitive to the voltage anomalies associated with $2\pi$ Josephson phase slips, which near $x$ =1/8 are observed both below and above $T_C$. These results document a new regime of nonlinear transport that shares features of fluctuating stripes and superconducting phase dynamics.**

The $c$-axis terahertz-frequency electrodynamics of high-$T_C$ cuprates [1] can be described by stacks of extended Josephson junctions [2,3] with distributed tunnelling inductance $L_J(x, y, t)$ between capacitively coupled planes ($x$ and $y$ are here the in-plane spatial coordinates and $t$ is the time). For low fields, $L_J = \frac{\hbar}{2eI_C}$ (where $e$ is the electron charge and $I_C$ the junction's critical current) is independent of space and time and a Josephson plasma resonance (JPR) is found at $\omega_{JPR} = 2\pi/\sqrt{L_J C}$ ($C$ is the equivalent capacitance of the planes). In most cuprates, $\omega_{JPR}$ ranges between gigahertz [4,5] and terahertz [6] frequencies. As characteristic for a plasmonic response, for $\omega > \omega_{JPR}$ the superconductor is transparent, corresponding to a positive dielectric function, $\varepsilon_1(\omega) > 0$, whereas for $\omega < \omega_{JPR}$ it is a perfect reflector with $R = 1$ and $\varepsilon_1(\omega) < 0$.

At high electric fields, the Josephson electrodynamics become nonlinear [7,8,9]. As radiation at photon energies below the average superconducting gap couples weakly to the order parameter amplitude, $|\psi(x, y, t)|$, the electrodynamics are primarily determined by changes of the order parameter phase. The phase difference between adjacent layers, $\Delta\varphi$ (henceforth referred to simply as $\varphi$), is therefore the relevant parameter, and the voltage that develops across the planes is expressed, according to the second Josephson equation, as $V = \frac{\hbar}{2e}\frac{\partial\varphi}{\partial t}$ (see Fig. 1(a)).

For an intuitive understanding of these physics, we consider here the case of a Josephson junction under a DC current bias, $I$ [10]. In this condition, the phase dynamics of the junction can be modelled as the motion of a fictitious particle in the "washboard" potential $U(\varphi) = -\frac{\hbar I_C}{2e}\left(cos\varphi + \frac{I}{I_C}\varphi\right)$, sketched in Fig. 1(b) [11]. These dynamics are well understood, both in the classical and in the quantum regime [12,13]. For $I \ll I_C$ the potential $U$ has local minima where the phase particle is trapped and oscillates at the JPR frequency (blue). An increase of $I$ has the effect of tilting the potential and decreasing the barrier between two neighbouring minima, progressively entering a nonlinear regime (red). For $I \sim I_C$ the phase "escapes" from the well and a net voltage develops at the junction's edges (green). By decreasing the bias current, the potential tilt is reduced and the particle will be retrapped in the new potential minimum [11].

A more comprehensive description of the effect of an intense transient THz field on the extended Josephson coupled cuprate planes (*i.e.*, with dimensions that exceed the London penetration depth) can be obtained by simulating the electrodynamics with the sine-Gordon equation [14,15], which in one dimension reads

$$\frac{\partial^2 \varphi(x,t)}{\partial t^2} + 2\gamma \frac{\partial \varphi(x,t)}{\partial t} + \omega_{JPR}^2 \sin\varphi(x,t) = \frac{c^2}{\varepsilon} \frac{\partial^2 \varphi(x,t)}{\partial x^2}.$$

Here, $x$ is the in-plane coordinate along the propagation direction of the pulse, $\varepsilon$ the relative dielectric permittivity, $c$ the speed of light in vacuum, and $\gamma$ a damping coefficient which accounts for the tunneling of normal quasiparticles.

Representative results of these simulations are shown in Fig. 1(c), where we report the time- and space-dependent Josephson phase in presence of an out-of-plane polarized driving field, with the same pulse shape shown in Fig. 1(a) and a spectrum peaked at $\omega_{JPR} = 0.5$ THz. The plots from left to right refer to increasing values of the peak field. For small electric fields and thus small phase, $\sin\varphi(x,t) \simeq \varphi(x,t)$ and the sine-Gordon equation yields a linear wave equation, leading to Josephson plasma waves (left panel). For progressively increasing fields, larger phase excursions are achieved and the nonlinear regime is accessed. Here, the oscillation frequency exhibits a redshift caused by the higher order terms in the $\sin\varphi$ expansion. Concomitantly, a rich phenomenology occurs, including odd harmonic generation and parametric amplification of Josephson plasma waves, which have been experimentally verified in a number of cuprate superconductors [16,17,18,19].

As the junction's current reaches the critical value, $I_C$, $2\pi$ phase slips are expected to develop across the junction, as already qualitatively captured by the analogy with the washboard potential depicted in Fig. 1(b). In the right panel of Fig. 1(c), the occurrence of a phase slip in the material is apparent.

It should be noted here that for the damping values, $\gamma$, and the shape of the THz pulse used for our experiments and included in these simulations, we do not expect soliton modes like those reported in Ref. [17]. Therein, these could be excited because the phase was driven with different THz pulse shapes, that is narrowband multi-cycle pulses from a free-electron laser.

As shown in Fig. 1(d)-1(f), evidence for phase slips is to be found if one can measure the junction's voltage $V = \frac{\hbar}{2e}\frac{\partial \varphi}{\partial t}$. Figure 1(e) shows the Fourier transform of the calculated Josephson voltage time profiles of Fig. 1(d), determined at the edge of the junction, *i.e.*, for $x = 0$. In the sample this point corresponds to the vacuum-material interface. As expected, the linear response exhibits a spectrum peaked at the driving frequency $\omega_{drive} \simeq \omega_{JPR}$. In the nonlinear regime (30 – 60 kV/cm) odd harmonics develop at $3\omega_{drive}$ and $5\omega_{drive}$, and these high frequency components grow dramatically when approaching the phase slip regime (160 kV/cm). A clear signature of the phase slip is evident in the peak voltage, determined at the fundamental frequency, $\omega_{drive}$, as a function of the driving electric field (Fig. 1(f)). Here, one observes an abrupt, discontinuous increase of the signal, which we shall take as a univocal fingerprint of the occurrence of a Josephson phase slip in the material.

We set out to measure the field scaling of the *c*-axis Josephson voltage, which develops at the surface of high-$T_C$ superconductors of the La$_{2-x}$Ba$_x$CuO$_4$ (LBCO) family, when driven by strong field single-cycle THz pulses. La$_{2-x}$Ba$_x$CuO$_4$ is an extensively studied "214" cuprate, exhibiting an anomalous suppression of the transition temperature, $T_C$, for doping levels near 1/8 (see schematic phase diagram in the inset of Fig. 2(c)) [20], related to the formation of "stripes", a peculiar charge- and spin-order pattern within the CuO$_2$ planes, consisting of one-dimensional chains of doped holes separated by antiferromagnetically ordered regions [21]. Recent studies [22] support the existence of a striped superfluid state at $T > T_C$ with a spatially modulated superconducting order parameter, a so-called pair-density-wave (PDW) state [23,24,25]. However, superfluid stripes are difficult to detect with conventional techniques (*e.g.*, scanning tunnelling microscopy), which are not sensitive to the order parameter phase, nor are they visible in linear *c*-axis optical measurements, due to their cross alignment in neighbouring CuO$_2$ layers (see *e.g.* inset of Fig. 3(c)) [24].

A recent experiment on LBCO showed that this frustration is removed in the nonlinear optical response [19]. A giant THz third harmonic, characteristic of nonlinear Josephson tunnelling, was observed in La$_{1.885}$Ba$_{0.115}$CuO$_4$ above the superconducting transition temperature, and up to the charge-ordering temperature, $T_{CO}$. Such response was modelled by assuming the presence of a PDW condensate, in which nonlinear mixing of optically silent tunnelling modes drives large dipole-carrying supercurrents.

Here, we make use of a new experimental probe technique to confirm the generation of odd harmonics of the phase and to measure voltage signatures of Josephson phase dynamics.

As in Ref. [19], we studied two compounds: La$_{1.905}$Ba$_{0.095}$CuO$_4$ (LBCO 9.5%), where the stripes are weaker and are only present below $T_{CO} \simeq T_{SO} \simeq T_C \simeq 32$ K (here $T_{CO}$ and $T_{SO}$ are the charge- and spin-ordering temperature, respectively), and La$_{1.885}$Ba$_{0.115}$CuO$_4$ (LBCO 11.5%), for which the superconducting transition is highly depleted ($T_C \simeq 13$ K) and the stripe phase is far more robust ($T_{CO} \simeq 53$ K and $T_{SO} \simeq 40$ K) [20].

The single-cycle THz pump pulses were generated in LiNbO$_3$ by the tilted-pulse front technique [26] with a spectrum peaked at ~0.5 THz, and were focused at normal incidence on the sample surface, with polarization along the out-of-plane crystallographic axis and maximum peak fields up to ~165 kV/cm [15].

The Electric Field Induced Second Harmonic (EFISH) [27] was sampled by 100-fs-long near-infrared (NIR) probe pulses with 800-nm wavelength, which were also polarized along the *c* axis, and scanned in time through the profile of the THz frequency pump. In equilibrium and without pump, no second harmonic was found.

The pump-induced 400-nm second harmonic intensity, $I_{SHG}$, generated at the surface was detected with a photomultiplier. By scanning the time delay between THz pump and NIR probe, the full second harmonic temporal profile could be determined.

The EFISH originates from a *four-wave* process involving the driving field, $E_{THz}$, and the near-infrared probe field, $E_{NIR}$. A third-order nonlinear polarization $P^{(3)}(\omega) = \chi^{(3)}(\omega_{NIR}, \omega_{NIR}, \omega_{THz})E_{NIR}E_{NIR}E_{THz}$ is generated at $2\omega_{NIR} \pm \omega_{THz}$, and measured by the detector as $I_{SHG} \propto |P^{(3)}(\omega)|^2$. In our setup, a passive second harmonic generated in the setup mixes with $P^{(3)}(\omega)$, allowing for *heterodyne* detection of a signal which is directly proportional to $P^{(3)}(\omega)$ [28].

As the NIR pulse duration is much shorter than the period of the THz field, and the frequency dependence of $\chi^{(3)}(\omega_{NIR}, \omega_{NIR}, \omega_{THz})$ within the THz spectral bandwidth can be neglected in first approximation, we are left with $I_{SHG} \propto E(t) = V(t)/d$. Here, $E(t)$ is the electric field that develops at the very surface of the material, within an interaction region of ~100 nm (set by the NIR probe penetration depth). In a Josephson junction, this quantity corresponds to $V(t)/d$, where $V(t)$ is the time-dependent voltage across the junction and $d$ the layer separation. Hence, our technique provides a direct experimental measurement of the Josephson voltage at the sample surface, enabling a one-to-one comparison with the calculated quantities in Fig. 1(e) and Fig. 1(f).

This type of measurement is different from that reported in Ref. [28]. Therein, coherent oscillations in the second harmonic signal, persisting over much longer time windows than the duration of the pump pulse, were attributed to an effective second-order susceptibility, which tracked the time-dependent oscillations of symmetry-odd, infrared-active modes. In our case, instead, we are dealing with an effect that is present only while the THz driving field excites the material, and therefore can be fully described in terms of an EFISH process.

In Fig. 2(a) we report the results of the experiment carried out in the superconducting phase at the lowest temperature ($T = 5$ K) in LBCO 9.5%. Therein, we show the Fourier transform of the measured EFISH time trace for various peak amplitudes of the THz driving field, having subtracted the same quantity measured at high temperature. The exact same type of measurements was also performed in the superconducting state of LBCO 11.5% (see Fig. S3(a) [15]).

In both compounds, the response was almost identical, characterized by a strong peak at the *fundamental* driving frequency, $\omega_{drive} \simeq 0.5$ THz (which in LBCO 9.5% overlaps with the equilibrium $\omega_{JPR}$ [29]), whose intensity scaled with the THz field amplitude. Concurrently, another contribution developed for $E_{peak} \gtrsim 100$ kV/cm, with oscillations appearing at the third harmonic frequency, $3\omega_{drive} \simeq 1.5$ THz and, only for the highest field data ($E_{peak} = 165$ kV/cm), signatures of a $5\omega_{drive} \simeq 2.5$ THz peak were also found. The latter, however, is comparable with the noise floor of our measurement and we refrain from analyzing it further, focusing on the $\omega_{drive}$ and $3\omega_{drive}$ terms.

The frequencies of these contributions did not depend strongly on $E_{peak}$. We found instead a systematically higher signal amplitude in LBCO 9.5% compared to LBCO 11.5%, likely related to the detuning of $\omega_{drive}$ compared to the resonance at $\omega_{JPR}$, which in the latter material is $\omega_{JPR} \simeq 0.2$ THz. Note that this frequency was contained in the drive spectrum but not at its peak (see Fig. S1 [15]).

Figures 2(b) and 2(c) report the $E_{peak}$ dependence in the EFISH signal of the $\omega_{drive}$ (b) and $3\omega_{drive}$ (c) contributions, estimated from multi-Gaussian fits to the data in Fig. 2(a) and Fig. S3(a) [15]. While the third harmonic displays a $\propto E_{peak}^3$ dependence, the fundamental peak shows a linear behavior only for low fields. A clear discontinuity is observed around 150 kV/cm in LBCO 9.5% and at ~130 kV/cm in LBCO 11.5%. This trend is reminiscent of that found in the voltage simulations of Fig. 1(f), and is indicative of Josephson phase slips in the superconducting phase of both materials when sufficiently high transient THz fields are applied. It is also not surprising that this effect is observed for slightly higher $E_{peak}$ values in LBCO 9.5%, a superconductor with higher critical temperature and larger phase rigidity.

The temperature dependence, which we report for both samples in Fig. 3(a) and Fig. 3(b) at a constant $E_{peak} \simeq 165$ kV/cm (see Fig. S4 [15] for the original spectra), provides additional information on these phenomena. While in the weakly-striped superconductor LBCO 9.5% the doubly-peaked EFISH response progressively reduces with increasing temperature approaching $T_C \simeq T_{SO} \simeq T_{CO}$, and then completely disappears in the stripe-free normal state, a much more striking effect is observed in LBCO 11.5%. Here, both the $\omega_{drive}$ and $3\omega_{drive}$ peaks clearly survive at $T > T_C$, all the way up to the charge-ordering temperature, $T_{CO}$.

The presence of phase slips in the stripe-ordered normal state of LBCO 11.5% is further underscored by the data in Fig. 3(c) and Fig. 3(d). These figures have the same structure as Fig. 2(b) and Fig. 2(c) (see Fig. S3(b) [15] for full spectra), with the difference that the data here were taken at $T = 30$ K, a temperature almost three times higher than $T_C$ in this compound. The response is virtually identical to that measured in the superconducting state, with a clear discontinuity in the field dependence of the fundamental peak (Fig. 3(c)) around ~130 kV/cm, a behavior which is very reminiscent of that predicted by the simulation of Fig. 1(f).

The observation of phase slips in the stripe-ordered state of $La_{1.885}Ba_{0.115}CuO_4$ complements the colossal third-harmonic signal previously reported in Ref. [19], providing new experimental evidence for interlayer coherence, and possibly for finite momentum condensation in the normal state of this cuprate. Whilst we are able to associate the voltage anomalies with phase slips in a finite momentum condensate, a comprehensive theory for this phenomenon has not yet been formulated.

A natural evolution for this field of study should address other forms of charge order that compete or coexist with superconductivity, such as those found in $YBa_2Cu_3O_{6+x}$ [30,31]. This same technique also has great potential to be applied in other regions of the cuprate phase diagram, where for example finite superfluid density and vanishing range phase correlations [32,33] are present, or other forms of density waves [34,35] have been discussed.

## Acknowledgments

The research leading to these results received funding from the European Research Council under the European Union's Seventh Framework Programme (FP7/2007-2013)/ERC Grant Agreement No. 319286 (QMAC). We acknowledge support from the Deutsche Forschungsgemeinschaft (DFG, German Research Foundation) via the excellence cluster 'The Hamburg Centre for Ultrafast Imaging' (EXC 1074 – project ID 194651731) and the priority program SFB925 (project ID 170620586). Work at Brookhaven is supported by the Office of Basic Energy Sciences, Division of Materials Sciences and Engineering, U.S. Department of Energy under Contract No. DE-SC0012704.

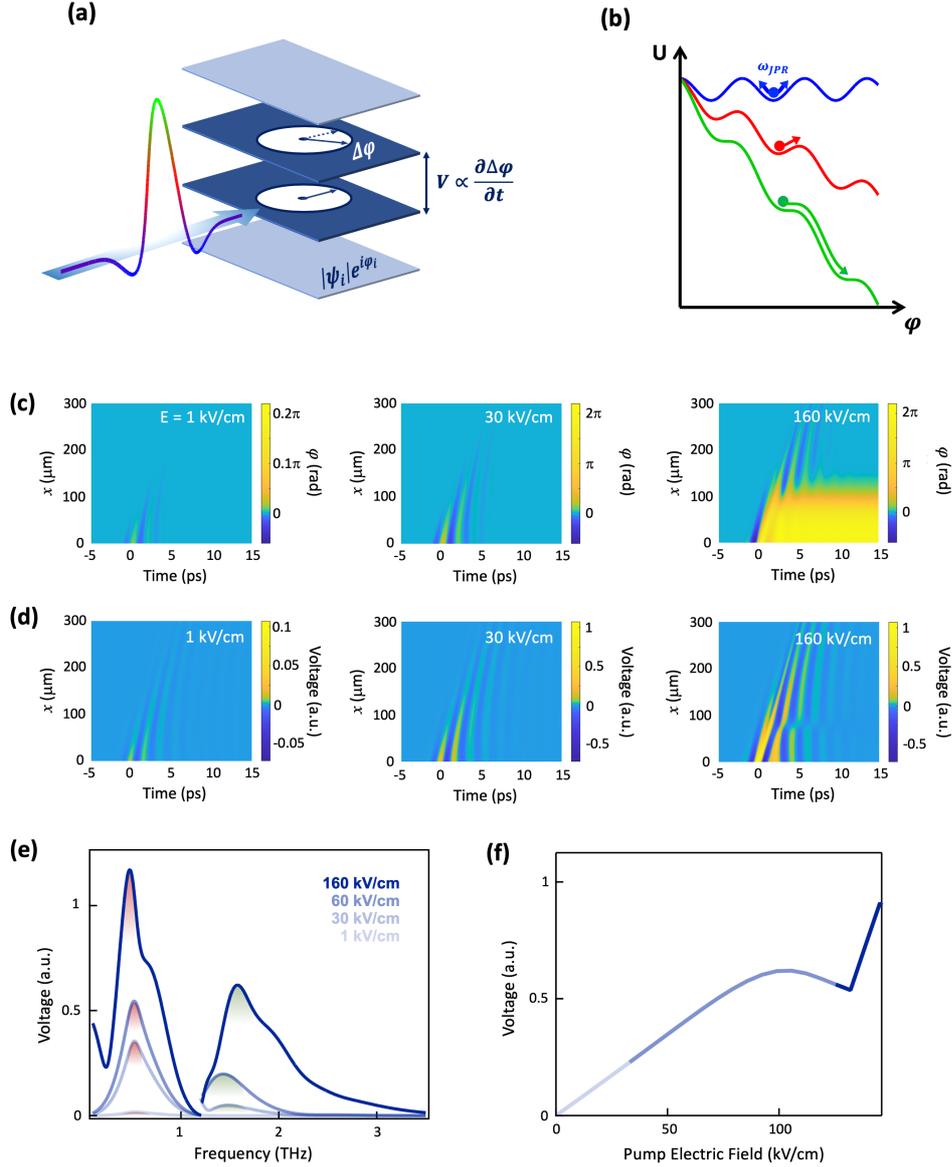

**Figure 1. (a)** Schematic representation of a layered superconductor in presence of a single-cycle THz driving field polarized along the out-of-plane direction. Depending on the instantaneous electric field carried by the pulse (color coded from blue through red to green), deformations of different magnitude in the space- and time-dependent order-parameter phase, $\varphi$, are achieved. According to the second Josephson equation, these phase gradients lead to the development of a voltage, $V$, across the layers. **(b)** Such scenario can be visualized in terms of a Josephson junction under the effect of a current bias, characterized by a tilted washboard potential. Each value of the electric field carried by the pulse corresponds to a certain tilting amplitude. Depending on the peak electric field, different regimes can be explored, starting from the linear one (blue), characterized by a response at the Josephson plasma frequency, $\omega_{JPR}$, to a nonlinear perturbative regime (red), until approaching the critical current (green), a condition in which one expects to observe phase slips. **(c)** Calculated space- and time-dependent interlayer phase, $\varphi(x,t)$, for a single extended Josephson junction illuminated by a THz pulse as that in (a), with a spectrum in resonance with $\omega_{JPR}$ and different THz peak field values, $E_{peak}$ = 1, 30, and 160 kV/cm. Here, $x$ denotes the in-plane direction along the pulse propagation. In the rightmost panel, the occurrence of a phase slip is apparent. **(d)** Corresponding space- and time-dependent interlayer voltages, $V(x,t) \propto \partial\varphi(x,t)/\partial t$. **(e)** Fourier transforms of the voltages in (d), calculated at the junction boundary ($x = 0$) for different $E_{peak}$ values. The peaks at $\sim\omega_{\text{drive}}$ and $\sim 3\omega_{\text{drive}}$ are shaded in red and green, respectively. All spectra at $\omega > 1.2$ THz have been multiplied by 30. **(f)** THz electric field dependent voltage, estimated at the peak frequency in (d) by scanning $E_{peak}$ continuously. A marked jump indicative of a phase slip is apparent.

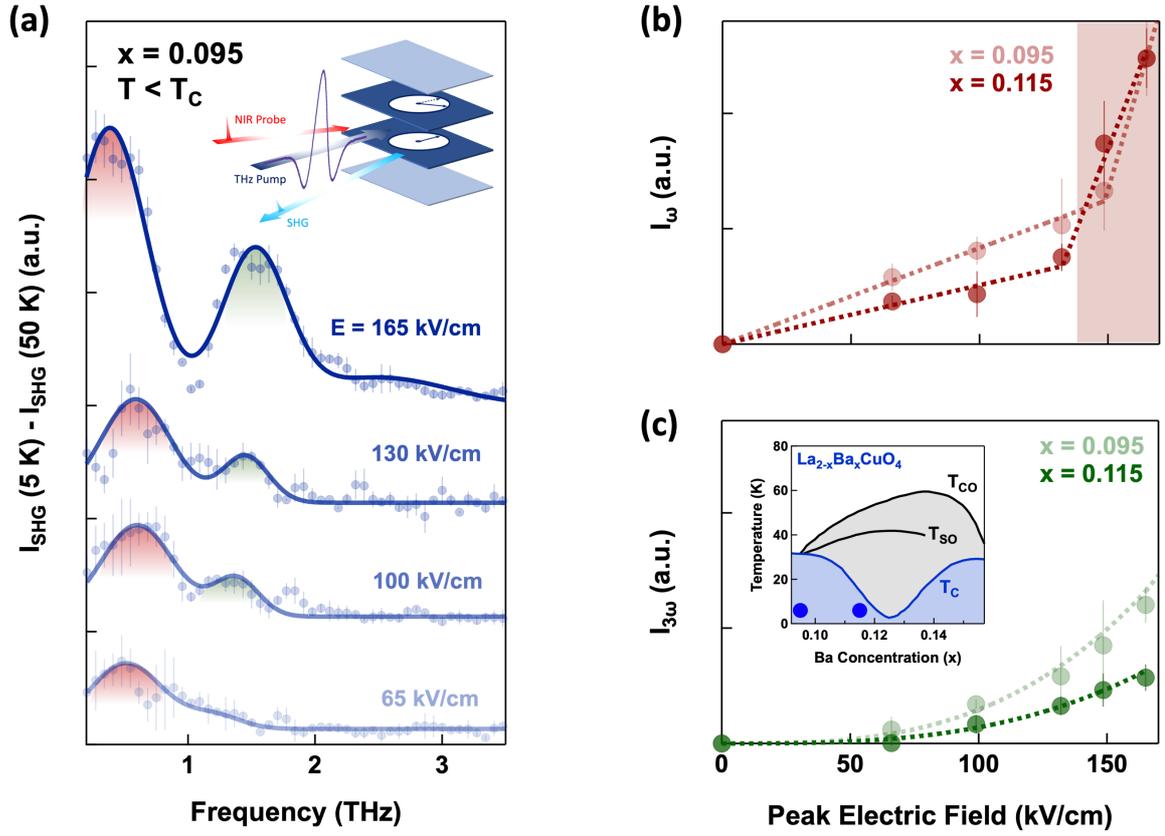

**Figure 2. (a)** Fourier transform of the second harmonic intensity measured in superconducting La$_{1.905}$Ba$_{0.095}$CuO$_4$ at $T = 5$ K $< T_C$ for different THz peak driving fields (see legend), after subtraction of the same quantity measured at $T > T_C$. The spectra have been vertically offsetted maintaining their relative amplitude. Uncertainty bars are standard errors estimated from different measurement sets. The peaks at $\sim\omega_{\text{drive}}$ and $\sim 3\omega_{\text{drive}}$ are shaded in red and green, respectively. Inset: Schematic of the experimental geometry, in which a single cycle THz pulse polarized along the $c$ axis is shone onto the sample surface together with a near-IR probe pulse (also polarized along $c$). The radiation generated at the second harmonic of the probe is then detected. The clocks in the layers represent amplitude (diameter) and phase (hand angle) of the superconducting order parameter. **(b)** Light red: Peak electric field dependence of the $\sim\omega_{\text{drive}}$ spectral component. Error bars are extracted here from the multi-Gaussian fits in (a). Dark red: Same quantity, measured in La$_{1.885}$Ba$_{0.115}$CuO$_4$ at $T = 5$ K $< T_C$, for which full sets of spectra are reported in Fig. S3(a) [15]. Dashed lines are linear guides to the eye, while the red shading highlights the phase slip regime. **(c)** Same quantities as in (b), extracted for the $\sim 3\omega_{\text{Drive}}$ component. Here the dashed lines are $\propto E^3$ fits. The vertical scales in (b) and (c) are mutually calibrated. Inset: Temperature-doping phase diagram of La$_{2-x}$Ba$_x$CuO$_4$, where the exact location of the investigated compounds is indicated by blue circles. Here, $T_{CO}$, $T_{SO}$, and $T_C$ are the charge-order, spin-order, and superconducting transition temperature, respectively.

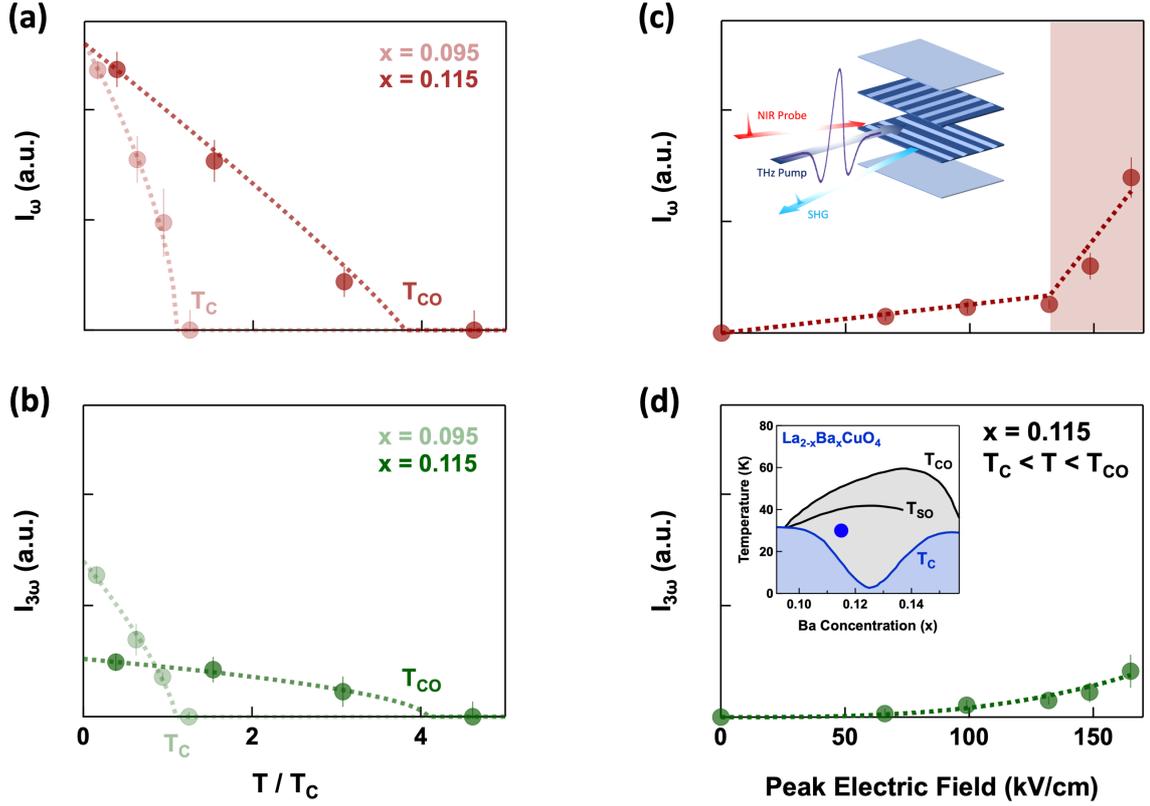

**Figure 3. (a)** Temperature dependence of the $\sim\omega_{\text{drive}}$ spectral component in the second harmonic intensity measured for both La$_{1.905}$Ba$_{0.095}$CuO$_4$ (light red) and La$_{1.885}$Ba$_{0.115}$CuO$_4$ (dark red), for a constant THz peak driving field of $\sim$165 kV/cm. Error bars are extracted from multi-Gaussian fits to the experimental spectra reported in Fig. S4 [15], while dashed lines are mean field fits. The horizontal temperature axis has been normalized by the superconducting $T_C$ of each compound. While for La$_{1.905}$Ba$_{0.095}$CuO$_4$ the response vanishes at $T_C = T_{CO} = 32$ K, in La$_{1.885}$Ba$_{0.115}$CuO$_4$ it persists well above $T_C = 13$ K, all the way up to $\sim T_{CO} = 53$ K. **(b)** Same quantities as in (a), extracted for the $\sim 3\omega_{\text{Drive}}$ component. **(c)** Peak electric field dependence of the $\sim\omega_{\text{drive}}$ component measured in La$_{1.885}$Ba$_{0.115}$CuO$_4$ at $T_C < T = 30$ K $< T_{SO}, T_{CO}$, for which full sets of spectra are reported in Fig. S3(b) [15]. Dashed lines are linear guides to the eye, while the red shading highlights the phase slip regime. Inset: Schematic of the experimental geometry. The charge order pattern in the layers is displayed as colored stripes. **(d)** Same quantity as in (c), extracted for the $\sim 3\omega_{\text{Drive}}$ component. Here the dashed lines are $\propto E^3$ fits. The vertical scales in (a), (b), (c), (d) are mutually calibrated. Inset: Temperature-doping phase diagram of La$_{2-x}$Ba$_x$CuO$_4$, where the exact location of the investigated compound is indicated by a blue circle.

# Terahertz phase slips in striped La$_{2-x}$Ba$_x$CuO$_4$

D. Fu[1], D. Nicoletti[1], M. Fechner[1], M. Buzzi[1], G. D. Gu[2], A. Cavalleri[1,3]*

[1] Max Planck Institute for the Structure and Dynamics of Matter, 22761 Hamburg, Germany

[2] Condensed Matter Physics and Materials Science Department, Brookhaven National Laboratory, Upton, NY, United States

[3] Department of Physics, Clarendon Laboratory, University of Oxford, Oxford OX1 3PU, United Kingdom

* e-mail: andrea.cavalleri@mpsd.mpg.de

# Supplemental Material

## S1. Simulation of the Josephson phase dynamics with the sine-Gordon equation

To simulate the dynamics of the Josephson junction we solved the one-dimensional sine-Gordon equation using a finite difference approach. Hereby, we assumed a single Josephson junction with semi-infinite layers, having the extended dimension along the $x$ direction. The evolution of the Josephson phase, $\varphi(x,t)$, is described by

$$\frac{\partial^2 \varphi(x,t)}{\partial t^2} + 2\gamma \frac{\partial \varphi(x,t)}{\partial t} + \omega_{JPR}^2 \sin\varphi(x,t) = \frac{c^2}{\varepsilon}\frac{\partial^2 \varphi(x,t)}{\partial x^2}. \qquad (S1)$$

Here, $\omega_{JPR}$ is the plasma frequency, $c$ the speed of light, $\varepsilon$ the relative dielectric permittivity, and $\gamma$ a damping factor accounting for the quasiparticle tunneling current. We used tabulated values for $\omega_{JPR}$ and $\varepsilon$, and set $\gamma$ as a fitting parameter.

We incorporated the THz driving field, $E(t)$, by setting the spatial and temporal phase evolution at the interface to the following condition:

$$\left.\frac{\partial \varphi(x,t)}{\partial t}\right|_{x=0} - c\left.\frac{\partial \varphi(x,t)}{\partial x}\right|_{x=0} = 2\omega_{JPR}\frac{E(t)}{E_c} \quad, \qquad (S2)$$

with $E_c = \Phi_0 \omega_{JPR}/(2\pi d)$, where $\Phi_0$ is the flux quantum and $d$ the distance between adjacent superconducting layers. Note that this is the same approach as that used in our previous works (see *e.g.*, Ref. [i]). For the simulations shown in Fig. 1 of the main text, we used the following values: $\omega_{JPR} = 0.53$ THz, $\varepsilon = 25$, $\gamma = 0.37$ THz, and $d = 10$ Å. The temporal and spatial grid have been tested for stability and convergence and finally set to $\Delta x = 1$ μm and $\Delta t = 4$ fs.

## S2. Experimental methods

Large single crystals of La$_{2-x}$Ba$_x$CuO$_4$ with $x = 0.095$ and $x = 0.115$ (~4 mm diameter), grown by transient solvent method, were studied here. These crystals belonged to the same batch of samples as reported in earlier works[i,ii,iii], and were cut and polished along the *ac* surface.

Laser pulses with 800-nm wavelength, ~100-fs duration and ~3.5 mJ energy were split into 2 parts (99% - 1%) with a beam splitter. The most intense beam was used to generate terahertz (THz) pulses by optical rectification in LiNbO$_3$ with the tilted pulse front technique[iv]. These pump pulses were collimated and focused at normal incidence onto the sample, and they were *s*-polarized (*i.e.*, perpendicular to the plane of incidence), corresponding to the crystallographic direction perpendicular to the CuO$_2$ planes (parallel to the *c* axis).

The THz beam spot diameter at the sample position was ~1.5 mm, with a maximum attainable field strength of ~165 kV/cm. The incident field strength was adjusted using a pair of wire grid polarizers. The peak electric field value at the sample position was estimated either by using a calibrated pyroelectric detector and accurately measuring the spot size, or by placing a GaP crystal in place of the sample itself and electro-optically sampling with a weak near-infrared gate beam. Both methods provided values in agreement within 20%.

An example time trace of the pump THz field, as well as the corresponding spectrum, are shown in Fig. S1. The spectrum is superimposed on the loss function of both samples measured in the superconducting state at the lowest temperature[i,v,vi]. This quantity shows a peak at the Josephson plasma frequency, providing a clear indication of the resonant pump condition, particularly for LBCO 9.5%.

The Electric Field Induced Second Harmonic (EFISH)[vii] was generated from 800-nm wavelength probe pulses (1% of the laser output). These were shone onto the sample with polarization along the *c* axis and scanned in time through the profile of the THz pump. In equilibrium and without pump, no second harmonic was found. The pump-induced 400-nm second harmonic intensity, $I_{SHG}(t)$, generated at the surface was detected with a photomultiplier. By scanning the time delay, $t$, between THz pump and NIR probe, the full second harmonic temporal profile could be determined.

In order to eliminate the effect of any slow drift in the signal, which could have originated for example from the accumulation of ice on the sample surface, we have undertaken a measurement procedure which involved the acquisition of scans at a given temperature interspersed with reference scans at a temperature higher than all relevant temperatures in the given compound. These reference scans were acquired at $T_{ref} = 50 \text{ K} > T_C, T_{SO}, T_{CO}$ for LBCO 9.5% and at $T_{ref} = 70 \text{ K} > T_C, T_{SO}, T_{CO}$ for LBCO 11.5%, after verifying in both materials that the EFISH signal was independent of temperature for all $T > T_{ref}$. Each measurement at these temperature pairs was then repeated several times, and the data were statistically analyzed to estimate the error bars shown, for example, in Fig. 2(a) of the main text.

An example of data processing on a single set of measurements is reported in Fig. S2. Therein, we show how time traces acquired at a certain $T < T_{ref}$ and at $T = T_{ref}$ are first subtracted in time domain and then Fourier transformed. Virtually identical results were obtained by performing the subtraction in frequency domain.

## S3. Additional data sets

In this section we report additional data sets to complement those shown in the main text. In particular, these are the original spectra from which we extracted, via multi-Gaussian fits, the dependence of $I_\omega$ and $I_{3\omega}$ on the peak driving THz field for LBCO 11.5% at $T = 5$ K (Fig. 2(b) and Fig. 2(c)) and $T = 30$ K (Fig. 3(c) and Fig. 3(d)), as well as the temperature dependences of the same quantities for both compounds (Fig. 3(a) and Fig. 3(b)).

In Fig. S3 we report THz field dependent spectra acquired in LBCO 11.5% at two temperatures ($T < T_C$ and $T_C < T < T_{CO}$), displaying a behavior virtually identical to that measured in the superconducting phase of LBCO 9.5% (Fig. 2(a) of the main text). Fig. S4 shows instead the variation of the second harmonic spectra as a function of temperature in both compounds, all acquired at the same peak THz field value of ~165 kV/cm. As discussed for Fig. 3(a) and Fig. 3(b) of the main text, while in LBCO 9.5% the response disappears at $T_C = 32$ K, in LBCO 11.5% it extends all the way to $T_{CO}$.

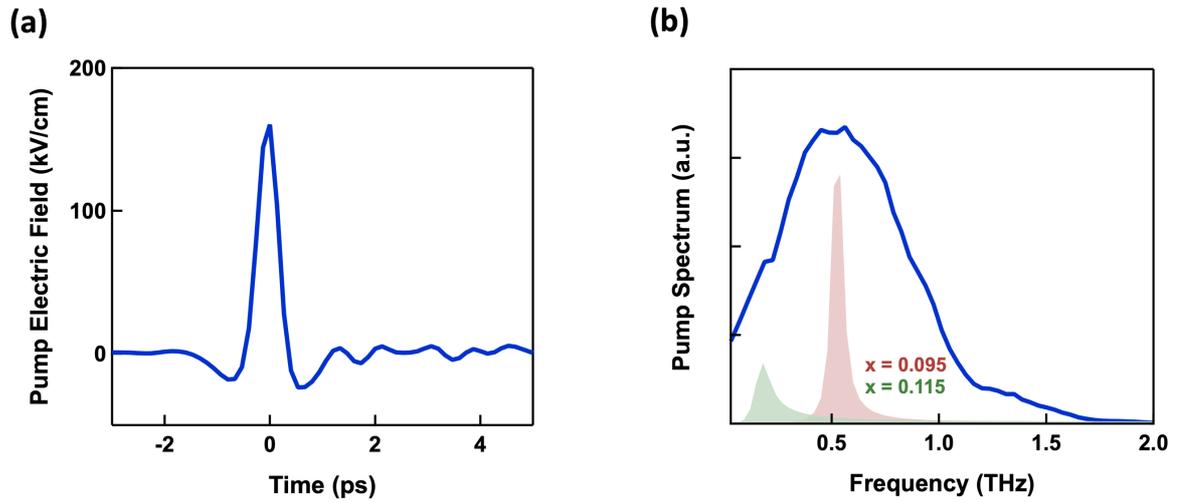

**Figure S1. (a)** Electric field profile of the THz pump pulses, measured at the sample position via electro-optic sampling in a 300-μm thick GaP crystal. **(b)** Blue curve: Corresponding Fourier transform spectrum of the time trace in (a). The red and green shadings indicate the energy loss function of $La_{1.905}Ba_{0.095}CuO_4$ and $La_{1.885}Ba_{0.115}CuO_4$, respectively, measured in the equilibrium superconducting state at $T = 5$ K $< T_C$. Notably, the Josephson plasmon of both compounds (peak frequency in the loss function) can be driven at (or close to) resonance by the THz pump field.

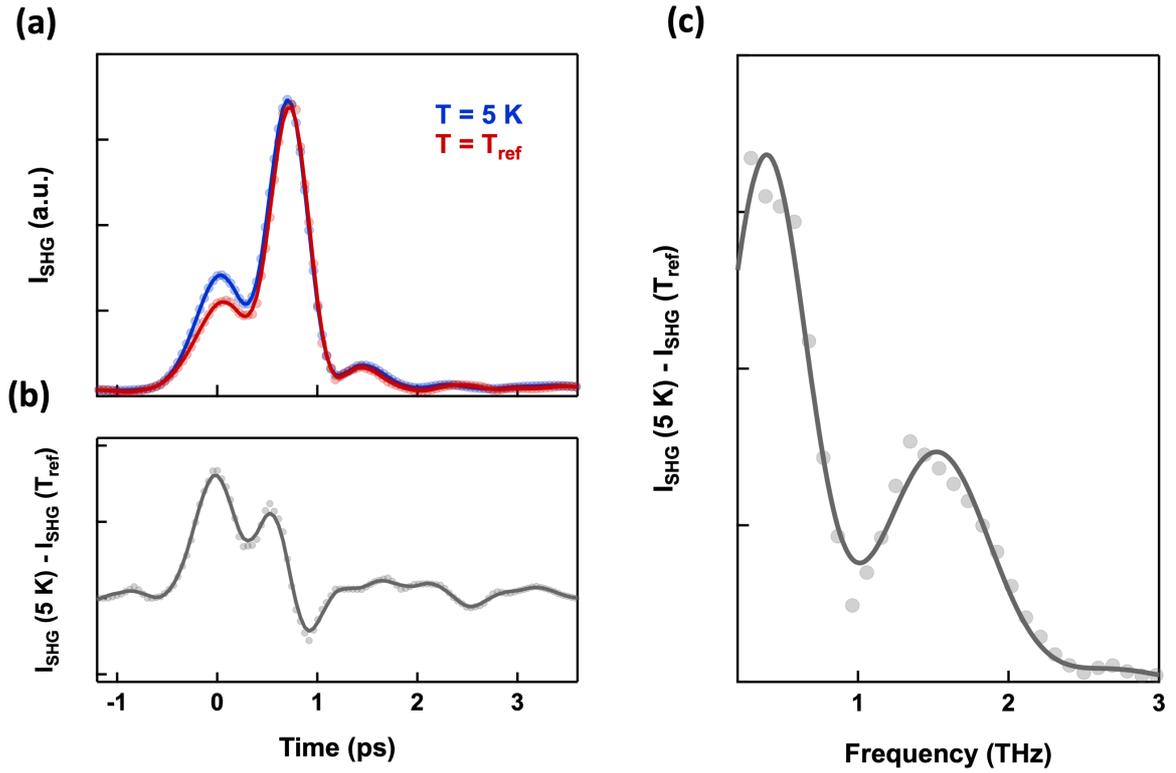

**Figure S2. (a)** Time-dependent Electric Field Induced Second Harmonic (EFISH) signal measured in $La_{1.905}Ba_{0.095}CuO_4$ at $T = 5$ K $< T_C$ (blue) and at the reference temperature $T_{ref} = 50$ K for a THz peak electric field of ~165 kV/cm. **(b)** Normalized EFISH signal, obtained by subtracting the two curves in (a) in time domain. **(c)** Frequency-dependent normalized EFISH response, obtained by Fourier transforming the curve in (b). This quantity is the same as that shown in Fig. 2(a) of the main text, as well as in Fig. S3 and Fig. S4 of the Supplemental Material.

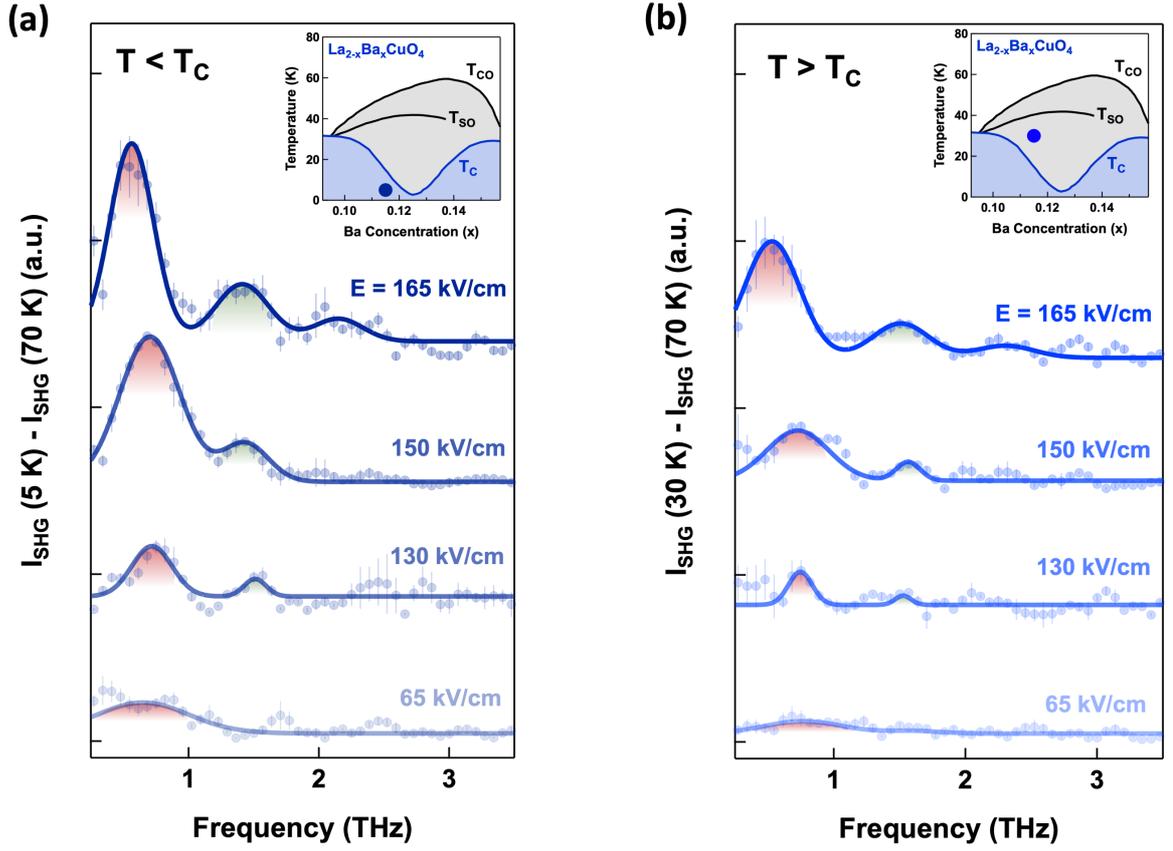

**Figure S3. (a)** Fourier transform of the second harmonic intensity measured in superconducting $La_{1.885}Ba_{0.115}CuO_4$ at $T = 5$ K $< T_C$ for different THz peak driving fields, after subtraction of the same quantity measured at $T > T_C, T_{SO}, T_{CO}$. The spectra have been vertically offsetted maintaining their relative amplitude. Uncertainty bars are standard errors estimated from different measurement sets. The peaks at $\sim\omega_{drive}$ and $\sim 3\omega_{drive}$ are shaded in red and green, respectively. Inset: Temperature-doping phase diagram of $La_{2-x}Ba_xCuO_4$, where the exact location of the investigated material is indicated by a blue circle. Here, $T_{CO}, T_{SO}$, and $T_C$ are the charge-order, spin-order, and superconducting transition temperature, respectively. **(b)** Same quantity as in (a), measured in $La_{1.885}Ba_{0.115}CuO_4$ at $T = 30$ K $> T_C$. Inset: Temperature-doping phase diagram as in (a).

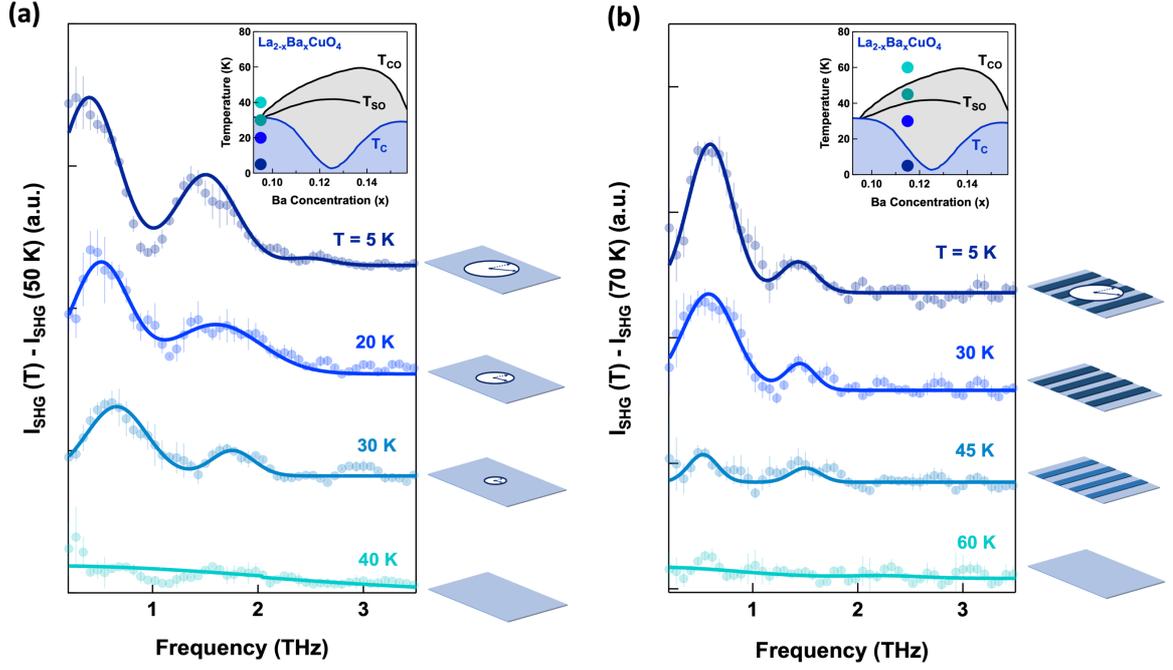

**Figure S4. (a)** Fourier transform of the second harmonic intensity measured in $La_{1.905}Ba_{0.095}CuO_4$ at different temperatures, for a peak THz field of ~165 kV/cm, after subtraction of the same quantity measured at $T = 50$ K $> T_C$. The spectra have been vertically offsetted maintaining their relative amplitude. Uncertainty bars are standard errors estimated from different measurement sets. Inset: Temperature-doping phase diagram of $La_{2-x}Ba_xCuO_4$, where the measured temperatures are indicated by circles ($T_{CO}$, $T_{SO}$, and $T_C$ are the charge-order, spin-order, and superconducting transition temperature, respectively). The graphics on the right represent the progressive reduction of the amplitude of the superconducting order parameter (clock diameter) with increasing temperature and its disappearance above $T_C$. **(b)** Same quantity as in (a) measured in $La_{1.885}Ba_{0.115}CuO_4$ at different temperatures across $T_C$, $T_{SO}$, and $T_{CO}$, for a ~165 kV/cm driving field. Normalization is done here by $T = 70$ K $> T_C, T_{SO}, T_{CO}$. Inset: Temperature-doping phase diagram as in (a). The graphics on the right represent the disappearance of the macroscopic superconducting order parameter (clock) when crossing $T_C$, as well as the gradual fainting of the stripe order, which coexists with superconductivity at $T < T_C$, and survives all the way up to $T = T_{CO}$.